\documentclass[a4paper]{jpconf}
\usepackage{graphicx}
\usepackage{hyperref}
\usepackage[noadjust]{cite}

\begin{document}
\title{Neutron stars: Observational diversity and evolution}

\author{S Safi-Harb}

\address{Department of Physics and Astronomy, University of Manitoba, Winnipeg MB R3T~2N2, Canada}

\ead{samar.safi-harb@umanitoba.ca}

\begin{abstract}
Ever since the discovery of the Crab and Vela pulsars in their
respective Supernova Remnants, our understanding of how neutron
stars manifest themselves observationally has been dramatically
shaped by the surge of discoveries and dedicated studies across
the electromagnetic spectrum, particularly in the high-energy
band. The growing diversity of neutron stars includes the highly
magnetized neutron stars (magnetars) and the Central Compact
Objects shining in X-rays and mostly lacking pulsar wind nebulae.
These two subclasses of high-energy objects, however, seem to be
characterized by anomalously high or anomalously low surface
magnetic fields (thus dubbed as `magnetars' and `anti-magnetars',
respectively), and have pulsar characteristic ages that are often
much offset from their associated SNRs' ages.
In addition, some neutron stars act `schizophrenic' in that they occasionally display properties that seem common to more than one of the defined subclasses.\\
I review the growing diversity of neutron stars from an
observational perspective, then highlight recent and on-going
theoretical and observational work attempting to address this
diversity, particularly in light of their magnetic field
evolution, energy loss mechanisms, and supernova progenitors'
studies.

\end{abstract}

\section{Introduction: Brief history of the neutron stars zoo}

This conference celebrates 50 years of neutron stars discovery.
Neutron Stars were discovered as Pulsars (PSRs) by Jocelyn Bell
and Antony Hewish in 1967 but predicted to exist back in 1934 by
Baade and Zwicky, just two years following the discovery of the
neutron by Chadwick. It is with the Crab and Vela pulsars
discovery in their respective supernova remnants (SNRs) that Baade
and Zwicky's 1934 prediction that supernovae (SNe) make neutron
stars was confirmed. The launch of imaging X-ray telescopes, such
as \textit{Einstein} (1980's),  followed by \textit{ROSAT} and
\textit{ASCA} (1990's) then by \textit{Chandra},
\textit{XMM-Newton} and \textit{Suzaku} (2000's), showed us how
these rotation-powered pulsars (RPPs) power synchrotron-dominated
pulsar wind nebulae (PWNe).

One of the neutron star subclasses introduced in the 1990's is the
`anomalous' X-ray pulsars (AXPs). It was recognized first by
Mereghetti \& Stella \cite{MereghettiStella(1995)} as a new class
of X-ray pulsars, since their spectra were much softer than those
of accretion-powered pulsars and had no binary companions;
furthermore their X-ray luminosity exceeds the energy available
from spin-down power, so they can not be powered by rotation.
Duncan \& Thompson \cite{DuncanThompson(1992)} suggested that
these are `magnetars', i.e. highly magnetized neutron stars with a
magnetic field exceeding the so-called `quantum critical field'
value of $B_{\rm QED}$~=~2$\pi~m_e^2
c^3$/$eh$~=~4.4$\times$10$^{13}$~G. AXPs are now merged with the
Soft Gamma-ray Repeaters (SGRs), as bursting high-energy sources
whose powerful outbursts are believed to be powered by their
magnetic field decay. Dubbed under the `Magnetars' family, these
objects are among the most extreme objects known in the universe
(see, e.g., \cite{Turollaetal2015}).

In the past decade, other sub-classes of neutron stars have
emerged, thanks to sensitive radio and X-ray observations. These
subclasses include the X-ray Dim Isolated Neutron Stars (XDINSs)
(or the Magnificent Seven \cite{Haberl2004}), the Rotating Ratio
Transients (RRATs \cite{McLaughlinetal2006}), the high-B radio
pulsars (HBPs) with $B$$\geq$$B_{\rm QED}$ \cite{NgKaspi2011}, and
the Central Compact Objects (CCOs), typified by the Cas~A CCO
\cite{pavlovetal2000,Gottheletal2013}. While we used to treat them
as different classes, we are now into the era of unifying these
sources thanks to a growing body of observational and theoretical
works. This review will highlight these works, focusing on neutron
stars with `extreme' magnetic fields (i.e. much smaller or higher
than the canonical value of $\sim 10^{12}$~G for the `classical'
rotation-powered pulsars) and stressing the fact that their
hosting SNRs  provide an independent means to address their
unification.

\begin{figure}
\begin{center}
\includegraphics[width=11cm]{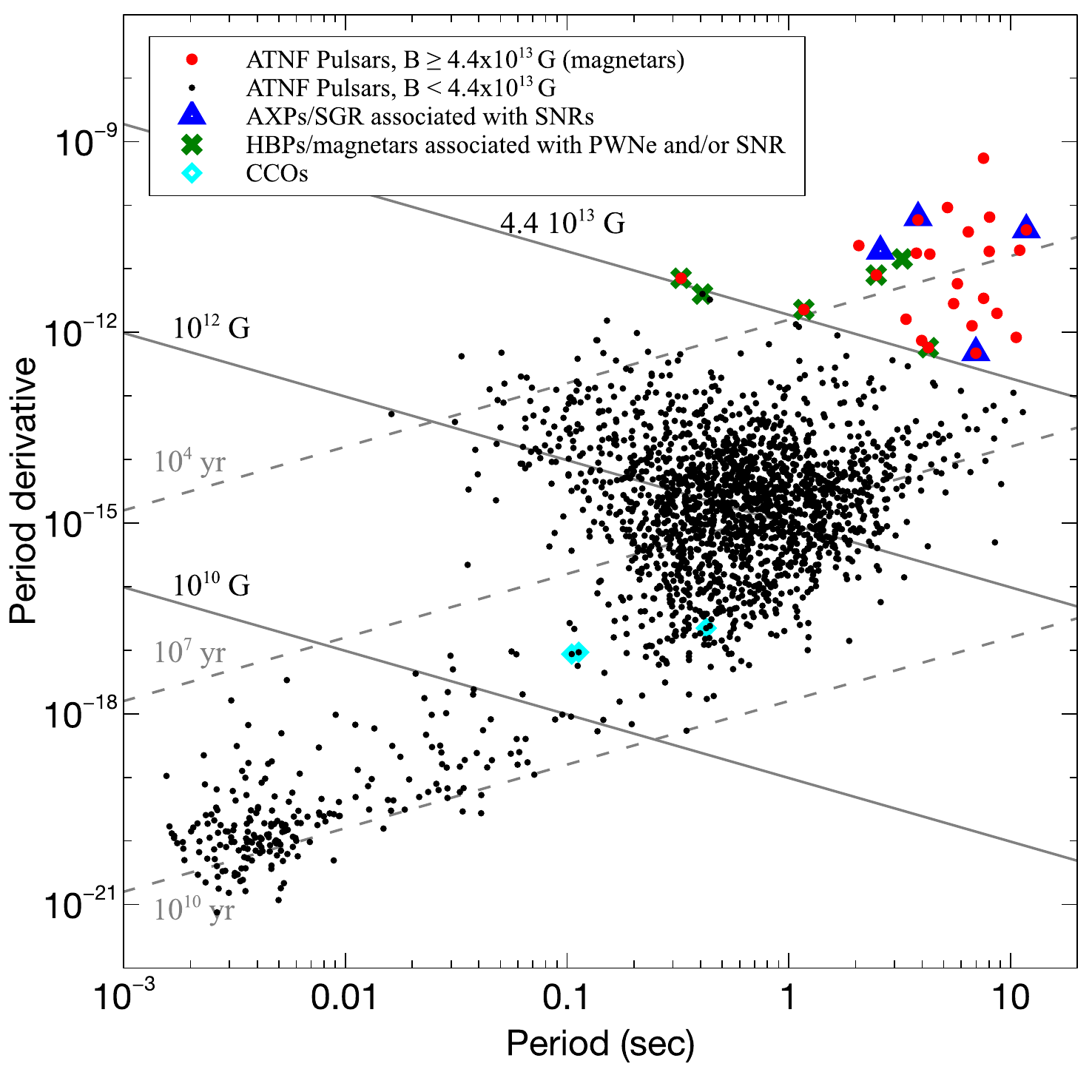}
\end{center}
\caption{\label{label}$P$--$\dot{P}$ diagram showing the diversity
of neutron stars. Lines of constant dipole magnetic field (solid
black) and characteristic age (dashed grey) are shown. The
4.4$\times$10$^{13}$~G line corresponds to the QED value of the
magnetic field that is traditionally used to separate magnetars
from the rotation-powered pulsars. The 3 pulsars shown in cyan
(diamond symbol) correspond to the 3 CCOs for which we have a
measured $P$ and $\dot{P}$. The 6 starred objects correspond to
the 5 HBPs and 1 magnetar associated with a PWN and/or SNR, with
the youngest and fastest two being PSR J1846--0258 in Kes~75 and
J1119--6127 in G292.2--0.5, both having now displayed
magnetar-like activity. The blue triangles correspond to the 3
AXPs + 1 SGR in our Galaxy known to be \textit{securely}
associated with an SNR \textit{with known age}.}
\end{figure}

\section{Pulsars Characteristics and diversity}

The diversity of pulsars is connected with their positions on the
so-called $P$--$\dot{P}$ diagram. The period $P$ and period
derivative $\dot{P}$ determine their spin-down energy
$\dot{E}$=$I$$\Omega$$\dot{\Omega}$ (where $I$ is their moment of
inertia and $\Omega$=2$\pi$/$P$), their surface dipole field
strength (at the magnetic equator)
$B$=3.2$\times$10$^{19}$~$\left(P\dot{P}\right)^{1/2}$~G, and
their characteristic age $\tau_c$=$P$/$2\dot{P}$. For some
pulsars, the braking index $n$=$\nu \ddot{\nu}/{\dot{\nu}}^2$
(where $\nu$=1/$P$) is measured (e.g, \cite{Espinoza2013}).

Figure~1 shows the $P$-$\dot{P}$ diagram for the 2,613 currently
known pulsars\footnote{
\url{http://www.atnf.csiro.au/research/pulsar/psrcat} (v1.56) },
revealing the growing diversity of neutron stars, and table~1
summarizes the secure PSR-SNR associations\footnote{
\url{http://www.physics.umanitoba.ca/snr/SNRcat} (SNRcat) } for
the pulsars with `extreme' magnetic fields. Below we briefly
comment on these growing subclasses of neutron stars, and then
address the blurring diversity and magnetic field evolution.

\subsection{ Magnetars and HBPs}

\begin{figure}
\begin{center}
\includegraphics[width=13.0cm]{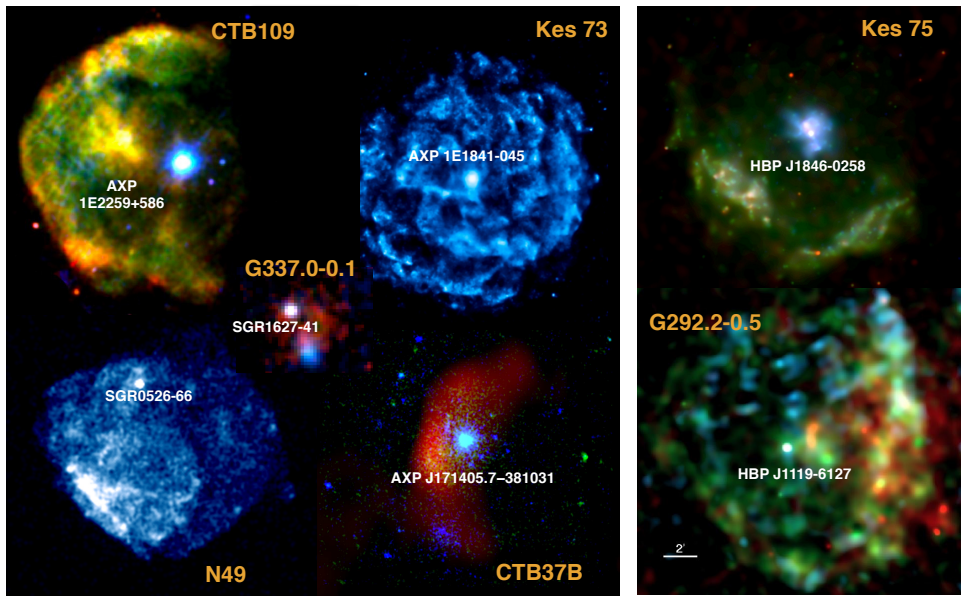}
\end{center}
{\vspace{-2cm}\caption{\label{label} Left: AXPs and SGRs securely
associated with SNRs. Right: HBPs associated with SNRs and having
displayed a magnetar-like behaviour. Figure parts \copyright\ AAS.
Reproduced with permission: CTB~109 (ESA/\textit{XMM-Newton}/M
Sasaki {\em et al} \cite{Sasakietal2004}), Kes~73 (NASA/CXC/U. of
Manitoba/H Kumar {\em et al}  \cite{Kumaretal2014}), N49
(NASA/CXC/Caltech/S Kulkarni {\em et al} \cite{Kulkarnietal2003}),
CTB~37B (MOST/ESA/\textit{XMM-Newton}/H Kumar \& S Safi-Harb),
G337.0--0.1 (ESA/\textit{XMM-Newton}/P~Esposito {\em et al}
\cite{Espositoetal2009}), Kes~75 (NASA/CXC/H Kumar \& S Safi-Harb
\cite{KumarSafi-Harb(2008)}), G292.2--0.5
(ESA/\textit{XMM-Newton}/H Kumar {\em et al}
\cite{Kumaretal2012}).}}
\end{figure}

There are currently 29 known magnetars (including 6
candidates)\footnote{
\url{http://www.physics.mcgill.ca/~pulsar/magnetar/main.html} }.
Magnetars are traditionally discovered as high-energy sources,
with no PWNe around them. Only 3 AXPs and 2 SGRs (one of which is
an extragalactic pulsar in the LMC SNR~N49) are securely
associated with SNRs (see figure~2). The High-B Pulsars (HBPs) are
a growing class of radio-detected pulsars with an inferred
magnetic field close to, or just above, the QED value. We know of
approximately a dozen such objects,
 and they are believed to be powered primarily by rotational energy.
The youngest HBPs, PSR J1846--0258 and J1119--6127 in the SNRs
Kes~75 and G292.2--0.5, respectively (figures~1 and 2), have
recently displayed magnetar-like activity providing conclusive
evidence for the connection between HBPs and magnetars (see
\S2.3).

\subsection{CCOs}
CCOs are X-ray emitting neutron stars found near SNR centres and
typified by the compact object
 in the SNR Cas~A (see figure~3).
The term CCO was coined by Pavlov {\em et al}
\cite{pavlovetal2000} following the first light \textit{Chandra}
discovery of the object.\footnote{In a footnote referring to the
Cas~A compact object, Pavlov {\em et al} \cite{pavlovetal2000}
state: `According to the convention recommended by the Chandra
Science Center, this source should be named CXO~J232327.9+584842.
We use the abbreviation CCO for brevity.'}

There are currently 14 CCOs\footnote{the total would be 15 CCOs if
we include the variable X-ray source 1E~161348--5055 in RCW~103.}
(including 6 candidates) known in our Galaxy
(\cite{Gottheletal2013, Pavlovetal2004}, SNRcat). These objects
are X-ray emitters with no optical or radio counterparts, and with
no evidence of PWNe surrounding them. X-ray pulsations have been
discovered from 3 CCOs (see table~1). Their timing properties
imply magnetic fields $\sim$$10^{10}$--$10^{11}$~G, much lower
than those of the traditional RPPs and magnetars, which led
Gotthelf \& Halpern \cite{GotthelfHalpern(2008)} to dub them
`anti-magnetars'. Spectroscopic studies support this
interpretation through the discovery of spectral line features
interpreted as cyclotron lines from a low $B$ (e.g.,
\cite{Sanwaletal2002,Bignamietal2003,DeLucaetal2012,GotthelfHalpern(2009)})
%[21--24])
or from modelling their thermal X-ray emission (e.g.,
\cite{HoHeinke2009,Suleimanovetal2014}).

These objects are also known to be quiet X-ray emitters (except
for the `CCO' in RCW~103) with their X-ray emission described by
two blackbodies and their X-ray luminosity (ranging from $\sim$ a
few$\times$10$^{32}$--10$^{34}$~erg~s$^{-1}$) exceeding their
spin-down energy. It is believed that their thermal spectra are
derived from residual cooling and partly from accretion of
supernova debris (e.g., \cite{GotthelfHalpern(2008)}). One of
their intriguing properties is their characteristic ages being
orders of magnitude greater than their hosting SNRs' ages (see
table~1 and figure~3).

\begin{figure}
{\vspace{-1cm} {\includegraphics[scale=0.5, bb= 0 40 600
600]{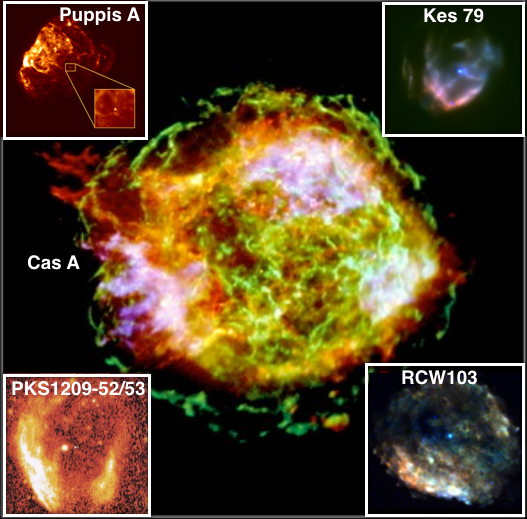} \hspace{-2pc}}}
\begin{minipage}[b]{15pc}\caption{\label{label}
The CCOs with known spin properties, in addition to Cas~A whose
CCO does not yet show pulsations (see table~1).
 Figure parts \copyright\ AAS. Reproduced with permission. Credits: Puppis~A: \textit{ROSAT}/R~Petre {\em et al} \cite{Petreetal1996}  Kes~79: NASA/CXC//P~Zhou {\em et al} \cite{Zhouetal2016}; PKS~1209--52/53: \textit{ROSAT}/Max-Planck-Institut f\"ur extraterrestrische Physik;
RCW~103: NASA/CXC/C~Braun \& S~Safi-Harb; Cas~A:
NASA/CXC/GSFC/U~Hwang {\em et al} \cite{Hwangetal2004}. }
\end{minipage}
\end{figure}

\subsection{Diversity blurred}
A growing body of observational and theoretical work has emerged
blurring this apparent diversity: (1) The discovery of
magnetar-like outbursts accompanied by spectral changes from the
rotation-powered HBP J1846--0258 in the SNR~Kes~75
\cite{KumarSafi-Harb(2008),Gavriiletal2008,Ngetal2008} and more
recently from the older HBP J1119--6127 in the SNR G292.2--0.5
\cite{Younesetal2016a,Kenneaetal2016,Gogusetal2016}
%[29--31].
(2) The discovery of transient radio emission from magnetars
(e.g., \cite{Camiloetal2006}) commonly discovered as high-energy
objects. (3) The discovery of low-$B$ magnetars, i.e. with
$B$$\leq$$B_{\rm QED}$, through their bursting activity
\cite{Reaetal2010,Scholzetal2012,Zhouetal2014}
%[33--35].
(4) The discovery of magnetar-like activity from 1E~161348--5055
\cite{DAiletal2016,Reaetal2016}, the variable central X-ray source
in the SNR RCW~103, classified by some authors as a CCO. (5) The
discovery of PWNe (which are characteristic of RPPs) around the
magnetar, Swift~J1834.9--0846 \cite{Younesetal2016b} and the HBP
J1119--6127
\cite{GonzalezSafi-Harb2003,Safi-HarbKumar2008,Blumeretal2017}
which, as mentioned above, recently revealed itself as a magnetar
-- see figure~1.

The thermal X-ray luminosities of some of the HBPs and magnetars
are systematically higher than those of the traditional radio
pulsars, suggesting that magnetic fields affect their X-ray
emission. Using 2D simulations of the fully-coupled evolution of
both temperature and magnetic field in neutron stars, Vigan\`o
{\em et al}  \cite{Viganoetal2013} unified the phenomenological
diversity of magnetars, HBPs and isolated nearby neutron stars by
varying their initial magnetic field, mass and envelope
composition. Furthermore, the magnetic field topology was argued
to play a key role in the observed properties of the bursting
activity of neutron stars. Perna \& Pons  \cite{PernaPons(2011)}
showed that the toroidal component, particularly its strength with
respect to the poloidal component, plays a significant role in the
frequency of the bursting activity and its dependence on the age
of the system. In addition, the role of fallback disks around
young isolated neutron stars has been highlighted by Alpar {\em et
al} \cite{Alparetal2013} to unify different classes of neutron
stars.

\section{SNRs associations shedding light on magnetic field evolution}

Generally it is assumed that neutron stars lose energy by spinning
down due to the emission of magneto-dipole radiation. However,
this simple model does not describe the neutron star population
for the following reasons. First, it predicts a braking index
$n=3$, which has not been observed in young neutron stars; most of
the measured indices are smaller than 3  (e.g.,
\cite{Espinoza2013}). Second, the pulsars securely associated with
SNRs show a remarkable disparity between their characteristic age
and the SNR age, in some cases differing by several orders of
magnitude (especially for the CCOs). Under the standard assumption
of constant magnetic field, we have: $P_0=P\left[ 1-(n-1)\tau
\frac{\dot{P}}{P} \right]^{\frac{1}{(n-1)}}$. Replacing $\tau$
with the SNR age (table~1), we are generally unable to explain the
observed braking indices and enforce the SNR's age to be equal to
the PSR's age with a constant $n$ (see, e.g., figure~2 in
\cite{RogersSafi-Harb2017}). Rogers \& Safi-Harb
\cite{RogersSafi-Harb2017,RogersSafi-Harb2016} addressed magnetic
field evolution focusing on the diverse population of neutron
stars with `anomalous' magnetic fields (i.e. much higher or lower
than the canonical value of 10$^{12}$~Gauss). These objects
include the AXPs, SGRs, HBPs and CCOs securely associated with
SNRs of known ages and listed in table~1.

\begin{table*}
\small
\begin{center}
\begin{tabular}{|c|c|c|c|c|}
\hline

Pulsar (SNR) & $P$  & $B$ ($\times$10$^{13}$~G) & $\tau$ (kyr) &
$\tau_{\rm SNR}$ (kyr) \\ \hline

AXP~1E~1841--045 (Kes~73)       &  11.79 & 70.3 &  $4.57$ & $0.75$--$2.10$ \\
AXP~1E~2259+586 (CTB~109)       &  6.98 & 5.9 & $230 $ & $10.0$--$16.0$ \\
CXOU~J171405.7--381031 (CTB~37B) &  3.83 & 50 & $0.95$ &
$0.35$--$3.15$ \\ \hline

SGR~0526--66 (N49)    & 8.05 & 56 & $3.36$ & $<4.80$ \\
SGR~1627--41 (G337.0--0.1) & 2.59  & 22 &  $2.2$ & $<5.0$ \\
\hline

HBP~J1846--0258 (Kes~75) &  0.3265 & 4.9 & 0.73 & $0.90$--$4.30$\\
HBP~J1119--6127 (G292.2--0.5) & 0.408 & 4.1 & 1.61 & $4.20$--$7.10$\\

\hline

RX~J0822.0--4300 (Puppis~A)       &  0.113 & 3.27$\times$10$^{-3}$ & 193              & $3.70$--$5.20$ \\
CXOU~J185238.6+004020 (Kes~79)    & 0.105 & 3.05$\times$10$^{-3}$ &  1.92$\times$10$^5$ & $5.40$--$7.50$ \\
1E~1207.4--5209 (PKS 1209--51/52) & 0.424 & 9.83$\times$10$^{-3}$ & 3.02$\times$10$^5$ & $2.00$--$20.0$ \\
 \hline

\end{tabular}
\caption{Pulsars with very high or very low magnetic field
securely associated with SNRs with known ages.} \label{table2}
\end{center}
\end{table*}

\subsection{Magnetic Field Decay}
Magnetic field decay has been invoked to describe the evolution of
AXPs and SGRs (e.g.,
\cite{Nakanoetal2015,DallOssoetal2012,Colpietale2000}). Magnetic
field decay channels depend on a variety of effects including
Ohmic dissipation, ambipolar diffusion and the Hall drift. While
for the low-$B$ pulsars with ages $\sim$10--100~kyr the Ohmic
dissipation is expected to dominate the magnetic field evolution,
for magnetar-strength-$B$ pulsars, the Hall effect provides the
dominant mechanism for the field evolution on timescales
comparable to their associated SNR ages.

\subsection{Magnetic field growth}
The hypothesis of fallback accretion goes back decades ago
\cite{Colgate1971,Zeldovichetal1972,Chevalier1989}. The
submergence of the magnetic field due to fallback accretion has
been explored by a number of authors (e.g.,
\cite{MuslimovPage1995,Geppertetal1999}), and revived recently
\cite{Ho2011,Ho2015,PopovTurolla2012,Bernaletal2010,Bernaletal2013,Igoshevetal2016,RogersSafi-Harb2016,RogersSafi-Harb2017,Eksi2017}
 for the following reasons:
(1) $n$$<$ 3 (see Section 3.3);  (2) timing of CCOs giving low
surface $B$$\sim$10$^{10}$--10$^{11}$~G; (3) spectroscopic
evidence for low surface magnetic field, as found, e.g., for the
CCOs in Cas~A, PKS~1209--51/52, and Puppis~A; (4) the highly
modulated pulsed signal in Kes~79's CCO  implying a much higher
internal magnetic field. This all supports growing evidence for
the submerged field scenario due to fallback accretion.

\subsection{Empirical model for magnetic field evolution}
In Rogers \& Safi-Harb
\cite{RogersSafi-Harb2016,RogersSafi-Harb2017}, both magnetic
field growth and decay are described within the same basic
framework, using empirical models for magnetic field evolution
developed for various classes of neutron stars. In this approach
the braking index is time-dependent: $n= 3 - 4 \tau
\dot{f_{\textit{j}}}/{f_{\textit{j}}}$, where the function
$f_{\textit{j}}(t)$ carries the time-dependence of the field.
Thus, field decay gives $n>3$ since $\dot{f_D}<0$, and field
growth gives $n<3$ since $\dot{f_G}>0$ . The sample shown in
table~1 (plus other RPPs with a measured braking index) gives
solutions to the field evolution in neutron stars,  some of which
are shown in figure~4 (see \cite{RogersSafi-Harb2017}  and
figure~5 therein for a more detailed and complete description of
the sample and fits considered). Magnetic field growth has been
used to fit the 3 CCOs (cyan), the 2 SGRs (red) and the
HBP~J1846--0258 in Kes~75 (green). AXPs have been fit with
magnetic field decay models highlighted by the grey area in
figure~4 \cite{Nakanoetal2015}. It is worth noting that the time
evolution of the CCOs' characteristic age explains the apparent
large discrepancy between the pulsars' ages (appearing very old)
and the ages of the associated young SNRs. In particular, for the
three systems shown, the PSR and SNR ages match at times
$\geq$10$^{4.5}$~yr, by which time the SNR would have mostly
dissipated. Therefore, the characteristic age for these pulsars
does not reflect their true age as long as they are within their
SNRs. This property, along with their inferred low asymptotic
field strength, lead to the suggestion that CCOs could be
ancestors of old isolated radio pulsars as long as they overcome
the accretion phase (which would explain their X-ray dominant
emission) and their surface field grows to the critical limit
required for radio emission.
 The late time evolution of the CCOs may also link them to the class of objects known as XDINS;
 radio-quiet X-ray pulsars with long periods
 and no apparent SNR associations, with some of
  these objects having magnetic fields $\geq$10$^{13}$~G, similar to the HBPs.

\begin{figure}[t]
\begin{center}
\begin{minipage}[h]{0.45\textwidth}
\hspace{-1.1cm}
\includegraphics[width=1.4\textwidth]
{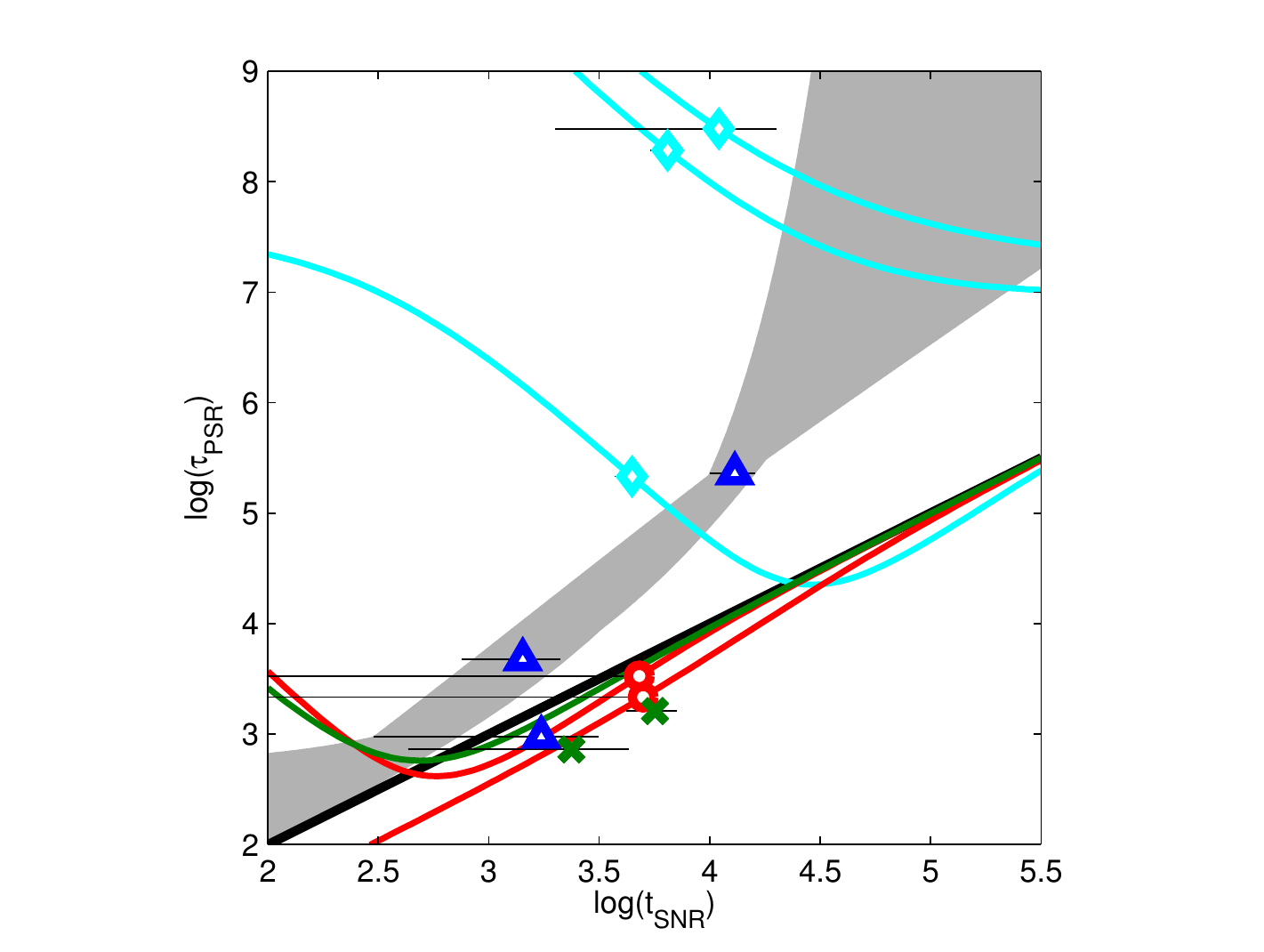}
\end{minipage}\hspace{0.05\textwidth}
\begin{minipage}[h]{0.45\textwidth}\caption{\label{label}
 Pulsar age versus SNR age shown with evolutionary tracks for evolving magnetic fields in PSR--SNR pairs.
The solid black diagonal line corresponds to equal PSR and SNR
ages. The colours/symbols match those used in figure~1, with the
cyan, red and green curves showing field growth fits to 3 CCOs, 2
SGRs and 1 HBP, respectively. See \S3.3 for details.}
\end{minipage}
\end{center}
\end{figure}

\section{SN progenitors}
The diversity of neutron stars can be related to the diversity in
SNe types resulting from core-collapse SNe (e.g.,
\cite{Chevalier2005,MilisavljevicFesen2017,PatnaudeBadenes2017}).

\subsection{Magnetar progenitors}
Which stars make magnetars? Two scenarios of interest have been
proposed in the literature. (1) The proto-neutron star model where
the neutron star is born with a few milliseconds period
\cite{DuncanThompson(1992)}, which predicts that the initial
kinetic energy of the supernova would be reflected in a
super-energetic hosting SNR. However, studies of magnetar SNRs and
HBP SNRs (albeit limited to a very small sample and subject to
low-resolution, CCD-type, X-ray spectroscopy) yield a `typical'
kinetic energy of the order of a
few$\times$10$^{50}$--10$^{51}$~ergs (e.g.,
\cite{Vink2008,Kumaretal2012}). Furthermore, the inferred ratio of
initial to current spin-period  ($P_0$/$P$) for the magnetars and
HBPs is not much smaller than 1, unlike what is predicted by the
proto-neutron star model. (2) The fossil field hypothesis where
magnetars are born from the most massive, most magnetic,
main-sequence stars \cite{FerrarioWickramasinghe2008}. Studies of
magnetar SNRs point towards massive progenitors
 (see, e.g., \cite{Safi-HarbKumar2013}),
 supporting the fossil field hypothesis.
However, this is not conclusive given that a few other studies
point to lower mass progenitors
\cite{BorkowskiReynolds2017,Daviesetal2009}.

\subsection{CCOs' progenitors}
Progenitor studies have been performed on the CCOs-hosting SNRs:
Cas~A, Puppis~A, Kes~79, RCW~103 and RX~J1713.7--3946. It is
interesting to note that these studies point to progenitor masses
around 20~$M_{\odot}$ (although inconclusively). Below we
summarize these studies.

\begin{itemize}
\item{Cas~A, the youngest historical type SNR in our Galaxy,
harbours the CCO CXOU~J232327.9+584842. It is commonly believed
that Cas~A results from a SN~IIb event with a progenitor mass of
$\sim$15--25~$M_{\odot}$ and may have lost its H envelope to
binary interaction (e.g., \cite{Youngetal2006}), although a
slightly higher mass (up to $\sim$30~$M_{\odot}$) progenitor has
been also discussed in the literature (e.g.,
\cite{Perez-Rendonetal2009}). } \item{Puppis~A is a
$\sim$4.5~kyr-old SNR harbouring the 112~ms CCO RX J0822--4300.
Spatially resolved spectroscopic studies of SNR ejecta using
\textit{Suzaku}, \textit{Chandra} and \textit{XMM-Newton}
 point to a 15--25~$M_{\odot}$ progenitor \cite{Katsudaetal2012,Hwangetal2008}}.
\item{ Kes~79 is a $\sim$5~kyr-old SNR harbouring the CCO
CXOU~J185238.6+004020 (PSR~J1852+0040).   A multi-wavelength
study, together with an X-ray spectroscopic study of the ejecta
compared to nucleosynthesis model yields, suggest a
15--20~$M_{\odot}$ progenitor \cite{Zhouetal2016})}.
\item{RX~J1713.7--3946 (G347.7--0.5) is a $\sim$1--2~kyr-old SNR
harbouring the CCO 1WGA J1713.4--3949. The measured metal
abundance ratios suggest that the progenitor star was a relatively
low-mass star ($\leq$20~$M_{\odot}$), However,  based on the
inferred blast wave velocity of $\sim$6,000~km/s which is
considered fast for such a
 core-collapse SNR, Katsuda {\em et al} \cite{Katsudaetal2015} propose that RX~J1713.7--3946 results from a SN Ib/c and that its progenitor is a member of an interacting binary. }
\item{RCW~103 is a  $\sim$2~kyr-old SNR hosting near its centre
the unusual CCO 1E~161348--5055 which had been recently proposed
to be a magnetar \cite{DAiletal2016,Reaetal2016}. Comparison of
the ejecta abundances inferred from X-ray spectra with supernova
nucleosynthesis yields models suggests a progenitor mass of
$\sim$18--20 $M_{\odot}$ \cite{Franketal2015}}.
\end{itemize}

\subsection{The Crab}

Going back to the `poster' outcome of a core-collapse SN, the
Crab's progenitor is known to be an $\sim$8--10~$M_{\odot}$
progenitor star. However, it has a low visible mass of
$\sim$5~$M_{\odot}$ and a small kinetic energy of $<$10$^{50}$~erg
for a young core-collapse SNR.
 Two scenarios have been discussed:
(1) a massive undetected shell beyond the visible PWN
\cite{Chevalier1977},
 (2) an electron-capture SN, an endothermic reaction of electrons captured in an O-Ne-Mg core of a super AGB star
 with a low energy ($\sim$10$^{50}$~erg) explosion \cite{Nomotoetal1982}.

This puzzle has been most recently addressed with the
\textit{Hitomi} X-ray satellite \cite{Takahashietal2014}. One of
the science goals for \textit{Hitomi} was to search for thermal
X-ray emission from the synchrotron-dominated SNRs, thanks to its
unprecedented spectral resolution and sensitivity to the thermal
X-ray emission with the Soft X-ray Spectrometer. A brief
observation took place just before the end of the \textit{Hitomi}
mission \cite{Hitomi2017}. The gate valve was still closed, which
hampered the sensitivity below 2 keV. While no thermal X-rays were
detected, as in previous dedicated searches (e.g., with
\textit{Chandra} \cite{Sewardetal2006}), a more constrained upper
limit on the X-ray emitting plasma has been established with
\textit{Hitomi} and past \textit{Chandra} and \textit{XMM-Newton}
studies. Furthermore, while a low-energy supernova explosion has
been favoured \cite{Smith2013,YangChevalier2015}, a higher energy
Fe core-collapse explosion could not be ruled out but implies,
depending on the environment, a very stringent upper limit on the
ambient density or mass loss rate  \cite{Hitomi2017}.

\section{Future prospects}
Despite significant advances in neutron stars physics in the past
50 years, many questions remain to be answered. We hope the
answers will be obtained with the planned X-ray missions.
\begin{itemize}
\item{The sample of PSR-SNR associations has been limited by
sensitivity and resolution. Furthermore, the X-ray emission from
CCO descendants and old radio pulsars would benefit from deep
X-ray observations and/or a sensitive X-ray survey. The upcoming
\textit{eRosita} X-ray satellite (Germany/Russia), expected to be
launched in 2018, will allow us to expand our sample and probe the
X-ray emission from the different classes of neutron stars.}
\item{To date, we do not have braking index measurement of
magnetars (SGRs and AXPs)  and other neutron stars subclasses, and
most measurements done to-date point to an index $n$$<$3.  Timing
studies of the different classes of neutron stars is needed to
shed light on their braking indices, which in turn addresses
magnetic field evolution. The recently launched \textit{ASTROSAT}
(India/Canada), \textit{NICER} (NASA), and the \textit{HXMT}
(China) X-ray satellites will provide a new window into timing
studies of pulsars.} \item{X-ray polarimetry is still lacking in
the field, but is crucial as it provides a direct measurement of
the neutron star's magnetic field and sheds light on its topology
(particularly for the magnetars and PWNe). \textit{XIPE} (ESA),
\textit{IXPE} (NASA) and  \textit{XTP} (China)  are currently
being planned for the near future.} \item{High-resolution X-ray
spectroscopy is needed to (a) provide an accurate measurement of
SNR ages, (b) probe SN progenitors and (c) provide a direct
measurement of the neutron star's magnetic field. Powerfully
demonstrated by the glimpse of the \textit{Hitomi} satellite on a
few targets before its loss, high-resolution X-ray spectroscopy is
now planned for the X-ray recovery mission (\textit{XARM},
JAXA/NASA; 2021 timescale) and in the more distant\
 future for  \textit{Athena} (ESA; 2028 timescale) and \textit{Lynx} (NASA; beyond 2030).}
\end{itemize}

\ack This review made use of NASA's Astrophysics Data System, the
U. of Manitoba's high-energy catalogue of SNRs (SNRcat
\cite{FerrandSafi-Harb2012}) and the ATNF PSR Catalogue
\cite{ATNF}. The model described in Section 3.3 is based on work
done with Adam Rogers. The author is supported by NSERC through
the Canada Research Chairs and the Discovery Grants Programs,  and
by MITACS and the Canadian Space Agency. The author thanks the
editors, particularly George Pavlov and Peter Shternin, and an anonymous referee for
helpful comments, and is grateful to the organizers for their
invitation and organization of an excellent conference.

\section*{References}
\providecommand{\newblock}{}

\end{document}